\documentclass[letter, scriptaddress, twocolumn,  prd, showpacs,showkeys]{revtex4-1}

	\usepackage{amsmath}
        \usepackage{mathrsfs}
	\usepackage{amssymb}
	\usepackage{graphicx} 
	\usepackage{makeidx}
	\usepackage{amsfonts}
	\usepackage[ansinew]{inputenc}
	\usepackage[usenames, dvipsnames]{pstricks}
	\usepackage{epsfig}
	\usepackage{pst-grad} 
	\usepackage{pst-plot} 
	\usepackage[colorlinks, hyperindex]{hyperref}
	\hypersetup
	{
		colorlinks, %
		citecolor=blue, %
		linkcolor=Green, %
		urlcolor=blue, %
	}

\makeindex

\begin{document}

\title{{\Large\bf Equivalence of Jordan and Einstein frames at the quantum level }}

\author{Sachin Pandey} 
\email{sp13ip016@iiserkol.ac.in}
\author{Narayan Banerjee}
\email{narayan@iiserkol.ac.in}
\affiliation{Department of Physical Sciences, \\
Indian Institute of Science Education and Research - Kolkata,\\
Mohanpur Campus, District Nadia,\\ 
West Bengal 741246, India.}

\begin{abstract}
It is shown that the Jordan frame and its conformally transformed version, the Einstein frame of nonminimally coupled theories of gravity, are actually equivalent at the quantum level. The example of the theory taken up is the Brans-Dicke theory, and the wave packet calculations are done for a homogeneous and isotropic cosmological model in the purest form of the theory, i.e., in the absence of any additional matter sector. The calculations are clean and exact, and the result obtained are unambiguous.
\end{abstract}
\maketitle
\section{Introduction}
A nonminimally coupled theory of gravity, where a field interferes with the curvature scalar, has two popular framework for its description. One is called the Jordan frame, where the theory is manifestly nonminimal in the sense that the interference term is visible in the action and also in the field equations derived from the action by means of a variational principle. In the second framework, known as the Einstein frame, the nonminimal coupling is broken by means of a conformal tranformation of the form ${\bar g_{\mu\nu}} = {\Omega}^2 g_{\mu\nu}$, the theory appears to be simpler and looks similar to General Relativity where the nonminimally coupled field appears as an additional ter in the matter sector. In Jordan frame, the Newtonian constant of gravity $G$ becomes a variable. Einstein's frame has a restored constancy of $G$ but the rest mass of the test particle becomes a function and thus one has to pay a bigger price, the validity of equivalence principle and hence the significance of the geodesic equation is lost. In fact this loss of the principle of equivalence is the key to understand the nature of the nonminimal coupling inspite of the apparent resemblance with general relativity.\\

The question of equivalence of these two frames for the same theory of gravity is yet to settled. The apparent mismatch of the two are quite obvious, Jordan frame description rests heavily on the principle of equivalence whereas the other does not respect that principle. The usual debate is centred around the question which frame is more dependable for the description of gravity. Cho indicated that the Einstein frame is more trustworthy for a physical description of gravity\cite{cho}. However, Faraoni and Gunzig\cite{faraoni} indicated that in the classical regime, Jordan frame is the reliable one on the consideration of gravitational waves. Chiba and Yamaguchi\cite{chiba} estimated various cosmological parameters in these frames and showed that they are different and hence indeed a matter of concern. \\

Some investigations, however, show that the apparent discrepancy in the results obtained in the two frames can actually be resolved. Salgado resolved the mismatch between the Cauchy problem in the two frames\cite{salgado}. Faraoni and Nadeau argued that the nonequivalence of the two frames actually comes down to a matter of interpretation, at least at the classical level\cite{faraoni2}.\\

Artymowski, Ma and Zhang\cite{ma} showed that Brans-Dicke theory looks different in the two framework in the context of loop quantum cosmology both in the presence or absence of another scalar field as the matter sector. \\

The question of equivalence is properly posed in the following way. The solutions obtained in one frame should be transformed into the second frame by means of the conformal transormation through which the metric components in the two versions are related and should then be compared with the solutions in the second frame. This would indicate whether the character of the frame itself introduces any feature which is not there in the other.\\

In the present work, we ask this question of equivalence at the quantum level. We work with Brans-Dicke theory\cite{brans}, easily the most talked about theory amongst the nonminimally coupled theories of gravity. In this theory, a scalar field $\phi$ is coupled with the Ricci scalar $R$ in the action. We work in a spatially flat, homogeneous and isotropic cosmological model in vacuum, and quantize the model following the standard canonical Wheeler-deWitt quantization scheme\cite{dewitt, wheeler}, and form the relevant wave packet $\Psi_{jordan}$ in Jordan frame. The action is then written in the Einstein frame via the conformal transformation  ${\bar g_{\mu\nu}} = \phi g_{\mu\nu}$ suggested by Dicke\cite{dicke}. We pretend that this is a completely different theory and quantize a same cosmological model following the same Wheeler-deWitt scheme. The wave packet $\Psi_{einstein}$ is formed. Naturally it looks different from the wave-packet in the Jordan frame. We now effect the inverse transformation in $\bar g_{\mu\nu}$ in the wave packet $\Psi_{einstein}$, and see that it is exactly the same as $\Psi_{jordan}$. The result is quite general, in the sense that this does not depend upon the parameter of the theory $\omega$. \\

In the next two sections the quantization of the cosmological model in vacuum is discussed in Jordan and Einstein frames respectively. Finally the result is critically analyzed in the last section.

\section{Jordan frame}
The relevant action in Brans Dicke theory without any contribution from the matter sector, in the so-called Jordan frame, is written as 
\begin{equation}
\label{action-jordan}
A=\int d^4 x \sqrt{-g} \bigg{[} \phi R + \frac{\omega}{\phi}\partial_\mu \phi \partial^\nu \phi \bigg{]},
\end{equation}

where $R$ is the Ricci scalar, $\phi$ is the scalar field and $\omega$ is a dimensionless parameter. It is generally believed that the higher the value of $\omega$, the closer the theory is to general relativity, and for $\omega \rightarrow \infty$, the two theories (GR and BD) are identical. However, it has been proved that this equivalence of the two theories is not at all generic\cite{soma}. \\

A spatially homogeneous and isotropic spacetime with a flat spatial section is given as
\begin{equation}
\label{metric}
ds^2 = n^2(t)dt^2 - a^2(t)(dr^2+r^2 d\theta^2 + r^2\sin ^2\theta d\phi^2),
\end{equation}
where the lapse function $n$ and the scale factor $a$ are functions of the time alone. With this metric, the Lagrangian can 
be extracted from the action (\ref{action-jordan}) as 
\begin{equation}
L=- \frac{6 \phi a \dot{a}^2}{n} -\frac{6 \dot{a} \dot{\phi}a^2}{n} + \frac{\omega}{n \phi}\dot{\phi}^2 a^3.
\end{equation}
With a change of variables as 
\begin{eqnarray}
\label{transf}
a(t)=e^{-\alpha/2 +\beta},\\
\phi(t) = e^{\alpha},
\end{eqnarray}

the Lagrangian can be written as
\begin{equation}
L =\frac{e^{-\alpha / 2 + 3 \beta}}{n} \bigg{[}-6\dot{\beta}^2+\frac{2\omega +3}{2}\dot{\alpha}^2 \bigg{]}.
\end{equation}
The corresponding Hamiltonian comes out to be 
\begin{equation}
H =  n e^{\alpha / 2 - 3 \beta}\bigg{[}-p_\beta^2+\frac{12}{2\omega+3}p_\alpha^2\bigg{]}.
\end{equation}

By a variation of the action in the first order with respect to the lapse function $n$, one has the Hamiltonian constraint as \\ 
\begin{center}
 $\mathcal{H}= e^{-\alpha / 2 + 3 \beta} H = 0$.
\end{center}
Now we consider a transformation of variables as $(\alpha, p_{\alpha})$ to $(T,p_T)$ given by  
\begin{eqnarray}
T=\frac{\alpha}{p_\alpha},\\
p_T=\frac{p_\alpha^2}{2}.
\end{eqnarray}
It is easily verified that $T$ and $p_T$ are canonically conjugate variables. \\
One can now write $\mathcal{H}$ as 
\begin{equation}
\mathcal{H}=-p_\beta^2+\frac{24}{2\omega+3}p_T.
\end{equation}
Here $\beta, T$ are the coordinates and $p_\beta, p_T$ are the corresponding canonically conjugate momenta. The canonical structure can be verified from the relevant Poisson brackets. \\

The Wheeler-deWitt (WDW) equation, $ \mathcal{H} \psi = 0$, can be written as
\begin{equation}
\label{wdejordan1}
\bigg{[}\frac{\partial^2}{\partial \beta^2} -i\frac{24}{2\omega+3} \frac{\partial}{\partial T}\bigg{]}\psi=0.
\end{equation}
The solution for above equation is obtained as
\begin{equation}
\label{psijordan1}
\psi_E(\beta, T)= e^{iET}\sin [\sqrt{24E/(2\omega+3)}\beta],
\end{equation}
or
\begin{equation}
\label{psijordan2}
\psi_E(a,\phi, T)= e^{iET}\sin [\sqrt{24E/(2\omega+3)}\ln (\sqrt{\phi}a)],
\end{equation}
where $E$ is a constant of separation.\\

Using $\int_0^{\infty}e^{-\gamma x}\sin {\sqrt{mx}}dx=\frac{\sqrt{\pi m}}{2 \gamma^{3/2}}e^{-m/4\gamma}$, wave-packet  can be written as
\begin{equation}
\label{wavepacket-jordan}
\Psi(a,\phi,T)= \sqrt{\frac{6\pi}{2\omega+3}}\frac{\ln (\sqrt{\phi}a)}{(\gamma-iT)^{3/2}} exp\bigg{[}-\frac{6 \ln^2 (\sqrt{\phi}a)}{(2\omega+3)(\gamma-iT)}\bigg{]}.
\end{equation}

\section{Einstein frame}
If one effects a conformal transformation given by 
\begin{equation}
\label{conf-trans}
\bar{g}_{\mu\nu} = \phi {g}_{\mu\nu}, 
\end{equation}

the action will look like

\begin{equation}
\label{action-einstein}
A=\int d^4x \sqrt{-\bar{g}}\bigg{[}\bar{R}+\frac{2\omega+3}{2}\partial_\mu \xi \partial^\nu \xi \bigg{]},
\end{equation}
where $\xi = ln\phi$ \cite{dicke}. 
The Lagrangian in this case can be written as 
\begin{equation}
L=-\frac{6\dot{\bar{a}}^2\bar{a}}{\bar{n}}+\frac{2\omega+3}{2\bar{n}}\dot{\xi}^2\bar{a}^3,
\end{equation}
and corresponding Hamiltonian becomes
\begin{equation}
H =(\bar{n} / \bar{a}^3) \bigg{[}-(\bar{a}^2/24)p_{\bar{a}}^2+\frac{1}{2(2\omega+3)}p_\xi^2\bigg{]}.
\end{equation}
The Hamoltonian constraint, as usual, can be obtained by varying the action with respect to the lapse function $\bar{n}$ as, 
$\mathcal{H}=\bar{a}^3H =0$. \\

Again with similar transformations as
\begin{eqnarray}
\bar{T}=\frac{\xi}{p_\xi},\\
p_{\bar{T}}=\frac{p_\xi^2}{2},
\end{eqnarray}
the WDW equation can be written as
\begin{equation}
\label{wdweinstein}
\bigg{[}\bar{a}^2\frac{\partial^2}{\partial \bar{a}^2} -i\frac{24}{2\omega+3} \frac{\partial}{\partial \bar{T}}\bigg{]}\bar{\psi}=0.
\end{equation}

One can easily see from the transformation equations that the scalar time parameters in the two frames, $T$ and $\bar{T}$ are actually equal.\\

With an operator ordering of the first term on the left hand side as $\bar{a}\frac{\partial}{\partial \bar{a}}\bar{a}\frac{\partial}{\partial \bar{a}}$, and taking $\chi = \ln \bar{a}$, the equation (\ref{wdweinstein}) can be written as 
\begin{equation}
\label{wdweinstein2}
 \bigg{[}\frac{{\partial}^{2}}{\partial {\chi}^{2}} - i \frac{24}{2\omega + 3}\frac{\partial}{\partial \bar{T}}\bigg{]} \bar{\psi} = 0.
\end{equation}

The solution for above equation can be given as
\begin{equation}
\label{psieinstein}
\bar{\psi}_E(\bar{a},\bar{T})=e^{iE\bar{T}}\sin [\sqrt{\frac{24E}{(2\omega+3)}}\ln (\bar{a})].
\end{equation}

The corresponding wave-packet is

\begin{equation}
\label{wavepacketeinstein}
\Psi(\bar{a},\bar{T})= \sqrt{\frac{6\pi}{2\omega+3}}\frac{\ln(\bar{a})}{(\gamma-i\bar{T})^{3/2}}exp\bigg{[}-\frac{6\ln^2(\bar{a})}{(2\omega+3)(\gamma-i\bar{T})}\bigg{]}.
\end{equation}

If one now revert the conformal transformation and go back to the Jordan frame, by using ${\bar a}^2 = a^2 \phi$ and $\xi = \ln \phi$ it is quite easy to see that the wave packet given in equation (\ref{wavepacketeinstein}) in the Einstein frame is exactly same as that in the Jordan frame given in equation (\ref{wavepacket-jordan}).

\section{Discussion}

The result obtained carries a clear message. If the action is not contaminated with other fields, such as a fluid, the Jordan and Einstein frames are completely equivalent in the sense that one can go from one description to the other at the final stage, i.e., at the level of the solution via the conformal transformation. This result is completely independent of the choice of the coupling constant $\omega$, which actually determines the deviation of the theory from general relativity. The work is carried out in Brans-Dicke theory.  Of course there are other more complicated nonminimally coupled theories where this has to be verified, but the message is clear.\\

Very recently a result contrary to this has been given\cite{barun1}, where it was shown that the wave packets in the two frames behave in different ways even after $\Psi_{einstein}$ is transformed back to the Jordan frame. The solutions were obtained for particular values of the Brans-Dicke parameter $\omega$, but that should not infringe upon the result. Perhaps the addition of a contribution of a fluid in the action results in the requirement of a proper ordering of operators, as the fluid variable and the geometry cannot be separated efficiently. \\

Furthermore, the conformal transformation, ${\bar g_{\mu\nu}} = {\Omega}^2 g_{\mu\nu}$, inflicts a change of units in the variables as indicated by Dicke\cite{dicke}, so one has to be careful about the interpretation of the results as shown by Faraoni and Nadeau\cite{faraoni2}. For various choices of units and their significance, we also refer to the early work by Morganstern\cite{morgan}. \\

It deserves mention that as the cosmic time $t$ is a coordinate and not a scalar parameter, the evolution of the quantum system requires a properly oriented scalar time parameter in the scheme, which is very efficiently constructed out of 
the fluid parameters as shown by Lapchinski and Rubakov\cite{rubakov}. In the present work, as no fluid is considered, the scalar time parameter ($T$ and $\bar T$ respectively in the two frames) is constructed from the scalar field and the scale factor following the work of Vakili\cite{vakili}. The derivatives with respect to $T$ and $\bar T$ appear in the first order in the Hamiltonian (equations (\ref{wdejordan1}) and (\ref{wdweinstein})) indicating their role in the scheme as time. One can easily check that the Poisson brackets $\{T, H\}$ and $\{\bar{T}, \bar{H}\}$ have the correct signatures which ensure the proper orientation of the time parameter. For a detailed description of this issue, we refer to the recent work of Pal and Banerjee\cite{sridip1}. \\

The present example is definitely a particular theory, namely the Brans-Dicke theory. But the calculations are so clean and the results are so unambiguous and general (independent of the Brans-Dicke coupling parameter $\omega$), that one can claim with confidence that the equivalence of the two frames are established, at the quantum level, at least when the action is taken in the pure form, i.e., without any matter field. In view of the recent result of the nonequivalence of the two frames in the presence of a fluid\cite{barun1}, one should perhaps look for a proper operator ordering which may yield this equivalence. If this attempt fails, then the emphasis should perhaps switch to the implication of the conformal transformation in the interpretation of matter variables. \\

Acknowledgement: The authors would thank Sridip Pal for some illuminating discussions. One of us (SP) is grateful to the CSIR, India for financial support.

\end{document}